\documentclass[english,aps,pra,twocolumn,showpacs]{revtex4}
\usepackage[T1]{fontenc}
\usepackage[latin9]{inputenc}
\usepackage{babel}
\usepackage{amsmath}
\usepackage{amssymb}
\usepackage{graphicx}
\usepackage[unicode=true]
 {hyperref}
\usepackage{breakurl}

\makeatletter

\newcommand{\noun}[1]{\textsc{#1}}

\@ifundefined{textcolor}{}
{%
 \definecolor{BLACK}{gray}{0}
 \definecolor{WHITE}{gray}{1}
 \definecolor{RED}{rgb}{1,0,0}
 \definecolor{GREEN}{rgb}{0,1,0}
 \definecolor{BLUE}{rgb}{0,0,1}
 \definecolor{CYAN}{cmyk}{1,0,0,0}
 \definecolor{MAGENTA}{cmyk}{0,1,0,0}
 \definecolor{YELLOW}{cmyk}{0,0,1,0}
 }

\makeatother

\begin{document}

\preprint{This line only printed with preprint option}

\title{Multipartite entanglement and high precision metrology}

\author{G\'eza T\'oth}

\email{toth@alumni.nd.edu}

\homepage{http://www.gtoth.eu}

\affiliation{Department of Theoretical Physics, The University of the Basque Country,
P.O. Box 644, E-48080 Bilbao, Spain}

\affiliation{IKERBASQUE, Basque Foundation for Science, E-48011 Bilbao, Spain}

\affiliation{Research Institute for Solid State Physics and Optics, Hungarian
Academy of Sciences, P.O. Box 49, H-1525 Budapest, Hungary\\
}

\pacs{03.67.Bg, 03.65.Ud, 42.50.St}
\begin{abstract}
We present several entanglement criteria in terms of the quantum Fisher
information that help to relate various forms of multipartite entanglement
to the sensitivity of phase estimation. We show that genuine multipartite
entanglement is necessary to reach the maximum sensitivity in some
very general metrological tasks using a two-arm linear interferometer.
We also show that it is needed to reach the maximum average sensitivity
in a certain combination of such metrological tasks. 
\end{abstract}

\date{\today}

\maketitle

\section{Introduction}

There has been a rapid development in the technology of quantum experiments
with photons \cite{BE04,KS05,PB00,KS07,WK09,PC09}, trapped ions \cite{SK00,HH05},
and cold atoms \cite{MG03}. In many of the experiments the goal is
to create a state with genuine multipartite entanglement \cite{BE04,KS05,PB00,SK00,HH05,KS07,WK09,PC09}.
From the operational point of view, the meaning of such an entanglement
is clear \cite{SK00,AB01}. An $N$-qubit quantum state state with
genuine $N$-partite entanglement cannot be prepared by mixing $N$-qubit
pure states, in which some groups of particles have not interacted.
Thus, the experiment presents something qualitatively new compared
to an $(N-1)$-qubit experiment. There is an extensive literature
on the detection of such entanglement (e.g., see Ref~\cite{GT09}
for a review.)

\begin{figure}
\includegraphics[width=8.6cm]{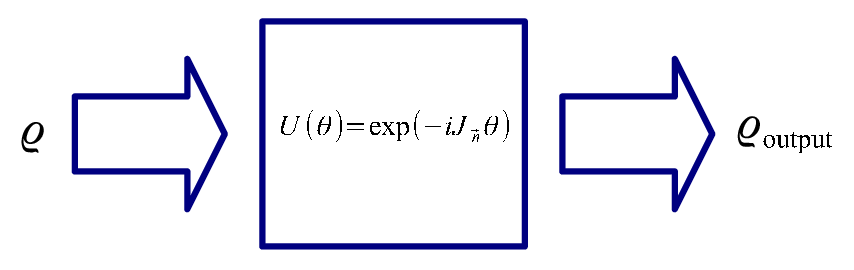}

\caption{(Color online) A basic problem of linear interferometry. The parameter
$\theta$ must be estimated by measuring $\varrho_{{\rm output}}.$
\label{fig:interfero}}
\end{figure}

One of the important applications of entangled multipartite quantum
states is sub-shotnoise metrology \cite{GS04}. In metrology, as can
be seen in Fig.~\ref{fig:interfero}, one of the basic tasks is phase
estimation connected to the unitary dynamics of a linear interferometer
\begin{equation}
\varrho_{{\rm output}}=e^{-i\theta J_{\vec{n}}}\varrho e^{+i\theta J_{\vec{n}}},
\end{equation}
where $\varrho$ is the input state of the interferometer, while $\varrho_{{\rm output}}$
is the output state, and $J_{\vec{n}}$ is a component of the collective
angular momentum in the direction $\vec{n}.$ The important question
is, how well we can estimate the small angle $\theta$ measuring $\varrho_{{\rm output}}.$
For such an interferometer the phase estimation sensitivity, assuming
\emph{any} type of measurement, is limited by the Quantum Cram\'er-Rao
bound as \cite{H76,H82}

\begin{equation}
\Delta\theta\ge\frac{1}{\sqrt{F_{Q}[\varrho,J_{\vec{n}}]}},
\end{equation}
where $F_{Q}$ is the quantum Fisher information. The relationship
between phase estimation sensitivity and entanglement in linear interferometers
has already been examined \cite{PS01}, and an entanglement condition
has been formulated with the sensitivity of the phase estimation,
that is, with the quantum Fisher information. It has been found that
some entangled states provide a better sensitivity in phase estimation
than separable states. It has also been proven that not all entangled
states are useful for phase estimation, at least in a linear interferometer
\cite{HG09}. Moreover, in another context, it has been noted that
multipartite entanglement, not only simple nonseparability is needed
for extreme spin squeezing \cite{SM01}. While this finding is not
directly related to the theory of the quantum Fisher information,
it does show that multipartite entanglement is needed for a large
sensitivity in certain concrete metrological tasks.

The question arises: Would it be possible to relate genuine multipartite
entanglement or any other type of multipartite entanglement to the
quantum Fisher information? Apart from the point of view of metrology,
this is also interesting from the point of view of entanglement criteria.
Bipartite entanglement criteria generalized for the multipartite case
typically detect any, that is, not necessarily genuine multipartite
entanglement. In fact, so far conditions for genuine multipartite
entanglement were mostly linear in operator expectation values (e.g.,
entanglement witnesses \cite{EntWit,TG10,JM11,BG11} or Bell inequalities
\cite{B64,M90,GB98,SU01,CN02}). There are also criteria quadratic
in operator expectation values \cite{U02,NK02,dVH11}, a strong criterion
based on the elements of the density matrix \cite{SG09,HM10} and,
some spin squeezing inequalities \cite{D11,T07,VH11}. For us, a starting
point can be that existing entanglement conditions based on the Wigner-Yanase
skew information \cite{C05} can also detect genuine multipartite
entanglement and many properties of the skew information and the quantum
Fisher information are similar.

In this paper, we examine what advantage states with various forms
of multipartite entanglement offer over separable states in metrology.
We show that in order to have the maximal sensitivity in certain metrological
tasks, $\varrho$ must be genuinely multipartite entangled. It can
also happen that an entangled state does not provide a sensitivity
for phase estimation larger than the sensitivity achievable by separable
states for any $J_{\vec{n}},$ however, the average sensitivity of
phase estimation is still larger than for separable states. Thus,
when asking about the advantage of entangled states over separable
ones in phase estimation, it is not sufficient to consider the sensitivity
in a single metrological task.

Now we are in a position to formulate our first main results; the
proofs are given later.

\textbf{\emph{Observation 1:}} For $N$-qubit separable states, the
values of $F_{Q}[\varrho,J_{l}]$ for $l=x,y,z$ are bounded as

\begin{equation}
\sum_{l=x,y,z}F_{Q}[\varrho,J_{l}]\le2N.\label{eq:F2ea}
\end{equation}
Here, $J_{l}=\tfrac{1}{2}\sum_{k=1}^{N}\sigma_{l}^{(k)}$ where $\sigma_{l}^{(k)}$
are the Pauli spin matrices for qubit $(k).$ Later we will also show
that Eq.~\eqref{eq:F2ea} is a condition for the average sensitivity
of the interferometer. All states violating Eq.~\eqref{eq:F2ea}
are entangled. Note that, according to Ref.~\cite{PS01} for separable
states we have 

\begin{equation}
F_{Q}[\varrho,J_{l}]\le N.\label{eq:F2eb}
\end{equation}

\textbf{\emph{Observation 2.}} For quantum states, the quantum Fisher
information is bounded by above as 

\begin{equation}
\sum_{l=x,y,z}F_{Q}[\varrho,J_{l}]\le N(N+2).\label{eq:all}
\end{equation}
Greenberger-Horne-Zeilinger states (GHZ states, \cite{GHZ}) and $N$-qubit
symmetric Dicke states with $\tfrac{N}{2}$ excitations saturate Eq.~\eqref{eq:all}.
Note that the above symmetric Dicke state has been investigated recently
due to its interesting entanglement properties \cite{KS07,T07,PC09}.
It has also been noted that the above Dicke state gives an almost
maximal phase measurement sensitivity in two orthogonal directions
\cite{HG09}. In general, pure symmetric states for which $\langle J_{l}\rangle=0$
for $l=x,y,z$ saturate Eq.~\eqref{eq:all}. 

Next, we consider $k$-producible states \cite{GT05,C05}. A pure
state is $k$-producible, if it is a tensor product of at most $k$-qubit
states \cite{GT05}. A mixed state is $k$-producible if it is a mixture
of pure $k$-producible states. 

\begin{figure}
\includegraphics[width=7cm]{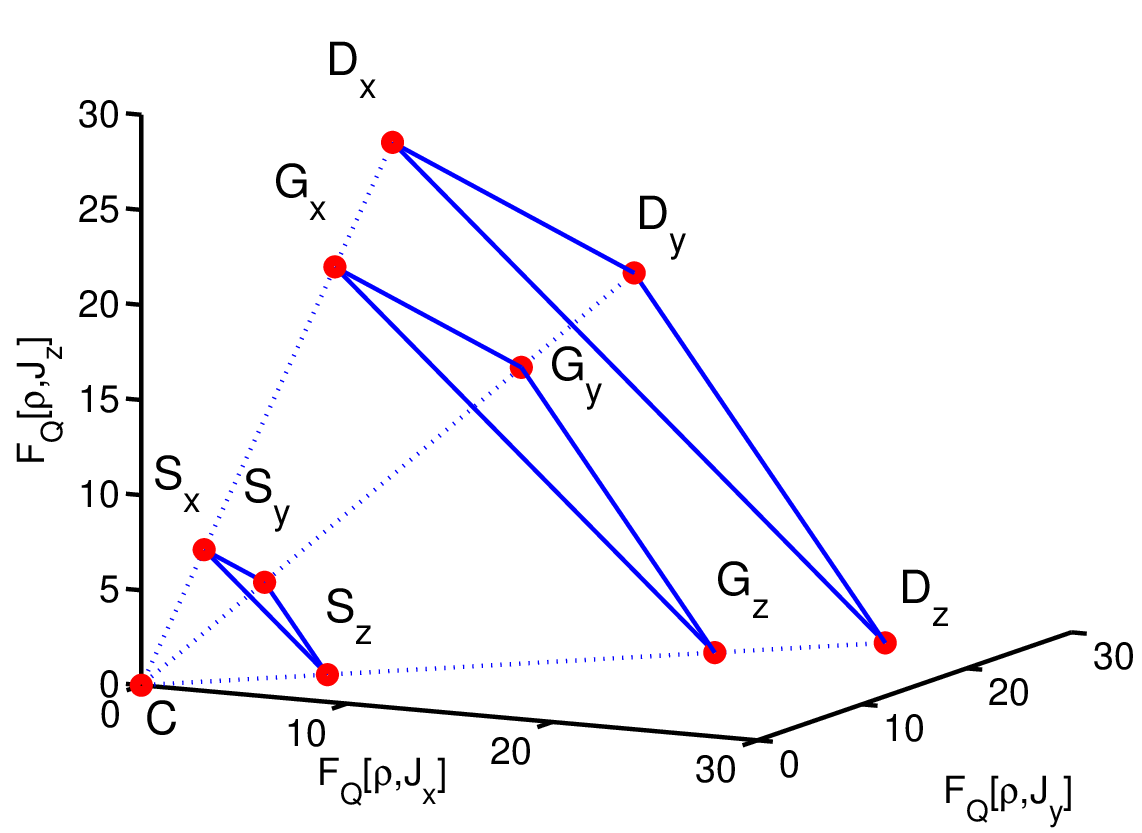}

\caption{(Color online) Interesting points in the $(F_{Q}[\varrho,J_{x}],F_{Q}[\varrho,J_{y}],F_{Q}[\varrho,J_{z}])$-space
for $N=6$ particles. Points corresponding to separable states satisfy
Eq.~\eqref{eq:F2ea} and are not above the $S_{x}-S_{y}-S_{z}$ plane.
Points corresponding to biseparable states satisfy Eq.~\eqref{eq:bisep}
and are not above the $G_{x}-G_{y}-G_{z}$ plane. All states corresponding
to points above the $G_{x}-G_{y}-G_{z}$ plane are genuine multipartite
entangled. For the coordinates of the $S_{l},G_{l},D_{l}$ and $C$
points, see Sec.~IV. \label{fig:Points}}
\end{figure}

\textbf{\emph{Observation 3.}} For $N$-qubit $k$-producible states
states, the quantum Fisher information is bounded from above by 

\begin{equation}
F_{Q}[\varrho,J_{l}]\le nk^{2}+(N-nk)^{2}.\label{eq:Nka-2}
\end{equation}
where $n$ is the integer part of $\tfrac{N}{k}.$ A condition similar
to Eq.~\eqref{eq:Nka-2} has appeared in Ref.~\cite{C05} for the
Wigner-Yanase skew information. 

\textbf{\emph{Observation 4.}} For $N$-qubit $k$-producible states
states, the sum of three Fisher information terms is bounded from
above by \cite{PH}

\begin{align}
\sum_{l=x,y,z}F_{Q}[\varrho,J_{l}]\le\;\;\;\;\;\;\;\;\;\;\;\;\;\;\;\;\;\;\;\;\;\;\;\;\;\;\;\;\;\;\;\;\;\;\;\;\;\;\;\;\;\;\;\;\;\;\;\;\;\;\;\;\;\;\;\;\;\;\;\;\;\;\;\;\nonumber \\
\begin{cases}
\begin{array}{c}
nk(k+2)+(N-nk)(N-nk+2)\\
nk(k+2)+2
\end{array} & \begin{array}{c}
\text{if }N-nk\ne1,\\
\text{if }N-nk=1.
\end{array}\end{cases}\label{eq:Nka}
\end{align}
where $n$ is the integer part of $\tfrac{N}{k}.$ Any state that
violates this bound is not $k$-producible and contains $(k+1)$-particle
entanglement.

Next, we consider criteria that show that the quantum state is not
biseparable. A pure state is biseparable if it can be written as a
tensor product of two multi-partite states \cite{AB01}. A mixed state
is biseparable if it can be written as a mixture of biseparable pure
states. The bounds for biseparable states for the left-hand-side of
Eqs.~\eqref{eq:Nka-2} and \eqref{eq:Nka} can be obtained from Observation
3 and 4 after taking $n=1$ and maximizing the bounds in those Observations
over $k=\lceil\frac{N}{2}\rceil,\lceil\frac{N}{2}\rceil+1,...,N-1,$
where $\lceil\frac{N}{2}\rceil$ is the smallest integer not smaller
than $\frac{N}{2}.$ Hence, we obtain \begin{subequations}

\begin{align}
F_{Q}[\varrho,J_{l}] & \le(N-1)^{2}+1,\label{eq:bisep-1}\\
\sum_{l=x,y,z}F_{Q}[\varrho,J_{l}] & \le N^{2}+1.\label{eq:bisep}
\end{align}
\end{subequations} Any state that violates Eq.~\eqref{eq:bisep-1}
or Eq.~\eqref{eq:bisep} is genuine multipartite entangled. 

The inequalities presented in Observations 1-3 correspond to planes
in the $(F_{Q}[\varrho,J_{x}],F_{Q}[\varrho,J_{y}],F_{Q}[\varrho,J_{z}])$-space
as can be seen in Fig.~\ref{fig:Points}. for $N=6$ particles. These
observations show that for fully separable states only a very small
fraction of the $(F_{Q}[\varrho,J_{x}],F_{Q}[\varrho,J_{y}],F_{Q}[\varrho,J_{z}])$-space
is allowed. This is also true for states with several forms of multipartite
entanglement, for example, $k$-producible states with $k\ll N.$
To reach the maximal phase sensitivity, genuine multipartite entanglement
is needed. 

The paper is organized as follows. In Sec. II, we prove Observations
1 and 2. In Sec. III, we prove Observation 3 and 4. In Sec. IV, we
examine the characteristics of the states corresponding to interesting
points in the $(F_{Q}[\varrho,J_{x}],F_{Q}[\varrho,J_{y}],F_{Q}[\varrho,J_{z}])$-space,
and determine which regions correspond to quantum states of different
degrees of entanglement. In Sec. V, we discuss some similarities to
entanglement detection with uncertainty relations. In Appendix A,
we present a unified framework to derive entanglement conditions independent
from the coordinate system chosen. In Appendix B, we give some details
of our calculations.

\section{Proof of Observations 1 and 2}

First, let us review some of the central notions concerning metrology
and the quantum Fisher information. For calculating many quantities,
it is sufficient to know the following two relations \cite{H76,H82,BC94,PS01}
for the quantum Fisher information
\begin{enumerate}
\item For a pure state $\varrho,$ we have $F[\varrho,J_{l}]=4(\Delta J_{l})_{\varrho}^{2}.$ 
\item $F[\varrho,J_{l}]$ is convex in the state, that is $F[p_{1}\varrho_{1}+p_{2}\varrho_{2},J_{l}]\le p_{1}F[\varrho_{1},J_{l}]+p_{2}F[\varrho_{2},J_{l}].$
\end{enumerate}
From these two statements, it also follows that $F[\varrho,J_{l}]\le4(\Delta J_{l})_{\varrho}^{2}.$ 

There is also an explicit formula for computing the quantum Fisher
information for a given state $\varrho$ and a collective observable
$J_{\vec{n}}$ for any $\vec{n}$ as \cite{HG09}

\begin{equation}
F_{Q}[\varrho,J_{\vec{n}}]=\vec{n}^{T}\Gamma_{C}\vec{n}.\label{eq:Fmat}
\end{equation}
Thus, the $\Gamma_{C}$ matrix carries all the information needed
to compute $F_{Q}[\varrho,J_{\vec{n}}]$ for any direction $\vec{n}.$
It is defined as \cite{HG09}

\begin{equation}
\left[\Gamma_{C}\right]_{ij}=2\sum_{l,m}(\lambda_{l}+\lambda_{m})\left(\frac{\lambda_{l}-\lambda_{m}}{\lambda_{l}+\lambda_{m}}\right)^{2}\langle l\vert J_{i}\vert m\rangle\langle m\vert J_{j}\vert l\rangle,\label{eq:GAMMAC}
\end{equation}
where the sum is over the terms for which $\lambda_{l}+\lambda_{m}\ne0,$
and the density matrix has the decomposition

\begin{equation}
\varrho=\sum_{k}\lambda_{k}\vert k\rangle\langle k\vert.
\end{equation}
Note that for pure states $[\Gamma_{C}]_{ij}=\langle J_{i}J_{j}+J_{j}J_{i}\rangle/2-\langle J_{i}\rangle\langle J_{j}\rangle$
\cite{HG09}. Later, we will present entanglement conditions with
$\Gamma_{C},$ besides entanglement conditions with $F[\varrho,J_{l}].$

\textbf{\emph{Proof of Observation 1.}}\emph{ }First, we show that
Observation 1 is true for pure states. We will use here the theory
of entanglement detection based on uncertainty relations \cite{uncent}.
According to this theory, for every $N$-qubit pure product state
of the form 

\textbf{\emph{
\begin{equation}
\vert\Psi_{{\rm P}}\rangle=\otimes_{n=1}^{N}\vert\Psi_{n}\rangle,
\end{equation}
}}the variance of the collective observable $J_{l}$ is the sum of
the variances of the single-qubit observables $j_{l}^{(n)}=\frac{1}{2}\sigma_{l}^{(n)}$
for the single-qubit states $\vert\Psi_{n}\rangle.$ Thus, we have
for the sum of the variances of the three angular momentum components
\cite{T04}

\begin{eqnarray*}
 &  & \sum_{l=x,y,z}(\Delta J_{l})_{\vert\Psi_{{\rm P}}\rangle}^{2}=\frac{1}{4}\sum_{l=x,y,z}\sum_{n=1}^{N}(\Delta\sigma_{l})_{\vert\Psi_{n}\rangle}^{2}=\\
 &  & \;\;\;\;\;\;\;\;\;\;\frac{1}{4}\sum_{n=1}^{N}(3-\langle\sigma_{x}^{(n)}\rangle^{2}-\langle\sigma_{y}^{(n)}\rangle^{2}-\langle\sigma_{z}^{(n)}\rangle^{2})=\frac{N}{2}.
\end{eqnarray*}
For the mixture of product states, that is, for mixed separable states,
Eq.~\eqref{eq:F2ea} follows from the convexity of the Fisher information.
\ensuremath{\hfill \Box}

Next, we show that Eq.~\eqref{eq:F2ea} can be interpreted as a condition
on the average sensitivity of the interferometer. First, note that
Eq.~\eqref{eq:F2ea} can be reformulated with the eigenvalues of
$\Gamma_{C}$ as

\begin{eqnarray}
{\rm Tr}(\Gamma_{C}) & \le & 2N.
\end{eqnarray}
Then, using Eq.~\eqref{eq:Fmat}, we obtain

\begin{equation}
{\rm avg}_{\vec{n}}(F_{Q}[\varrho,J_{\vec{n}}])={\rm avg}_{\vec{n}}\{{\rm Tr}[\Gamma_{C}(\vec{n}\vec{n}^{T})]\}=\left\{ \right\} {\rm Tr}(\Gamma_{C}\frac{\openone}{3}),
\end{equation}
where averaging is over all three-dimensional unit vectors. Thus,
Eq.~\eqref{eq:F2ea} can be rewritten as a condition for the average
sensitivity of the interferometer as

\begin{eqnarray}
{\rm avg}_{\vec{n}}(F_{Q}[\varrho,J_{\vec{n}}]) & \le & \frac{2}{3}N.
\end{eqnarray}

Let us calculate now the maximum of the left-hand side of Eq.~\eqref{eq:F2ea}. 

\textbf{\emph{Proof of Observation 2.}} We have to use that the quantum
Fisher is never larger than the corresponding variance,

\textbf{\emph{
\begin{equation}
\sum_{l=x,y,z}F(\varrho,J_{l})\le4\sum_{l=x,y,z}(\Delta J_{l})^{2}\label{eq:ineq}
\end{equation}
}}and that the sum of the variances are bounded from above\textbf{\emph{
\begin{equation}
4\sum_{l=x,y,z}(\Delta J_{l})^{2}\le4\sum_{l=x,y,z}\langle J_{l}^{2}\rangle\le N(N+2).\label{eq:ineq-1}
\end{equation}
}}For pure states, Eq.~\eqref{eq:ineq} is saturated. The second
inequality of Eq.~\eqref{eq:ineq-1} appears as a fundamental equation
in the theory of angular momentum. For symmetric states with $\langle J_{l}\rangle=0$
for $l=x,y,z$, both inequalities of Eq.~\eqref{eq:ineq-1} are saturated.
Hence GHZ states and Dicke states with $\tfrac{N}{2}$ excitations
saturate Eq.~\eqref{eq:all}. \ensuremath{\hfill \Box}

\section{Bounds for multipartite entanglement\emph{ }}

In this section, we present the proof of Observation 3 and 4 and also
compute some bounds for other types of multipartite entanglement.
For that, we use ideas similar to the ones in the proof of Observation
1. 

\textbf{\emph{Proof of Observation 3.}} Let us consider pure states
that are the tensor product of at most $k$-qubit entangled states\textbf{\emph{
\begin{equation}
\vert\Psi_{k-{\rm producible}}\rangle=\vert\Psi_{1}^{(N_{1})}\rangle\otimes\vert\Psi_{2}^{(N_{2})}\rangle\otimes\vert\Psi_{3}^{(N_{3})}\rangle\otimes\vert\Psi_{4}^{(N_{4})}\rangle\otimes.....,\label{eq:ttt-2}
\end{equation}
}}where $N_{m}\le k$ is the number of qubits for the $m^{th}$ term
in the product. Hence, based on using $(\Delta J_{l})_{\vert\Psi_{m}^{(N_{m})}\rangle}^{2}\le\frac{N_{m}^{2}}{4}$
for the $N_{m}$-qubit units, we obtain

\begin{eqnarray*}
(\Delta J_{l})_{\vert\Psi_{k-{\rm producible}}\rangle}^{2} & = & \sum_{m}(\Delta J_{l})_{\vert\Psi_{m}^{(N_{m})}\rangle}^{2}\le\sum_{m}\frac{N_{m}^{2}}{4}.
\end{eqnarray*}
For the case when $k$ is a divisor of $N,$ the largest variance
can be obtained for a state for which all $N_{m}=k.$ Hence, for the
state Eq.~\eqref{eq:ttt-2} we obtain 
\begin{eqnarray}
(\Delta J_{l})^{2} & \le & \frac{N}{k}\times\frac{k^{2}}{4}.\label{eq:kk2-1}
\end{eqnarray}
If $k$ is not a divisor of $N$ then at least one of the states in
the tensor product of Eq.~\eqref{eq:ttt-2} will have fewer than
$k$ qubits. The maximum for the sum of the variances is obtained
if all but a single state has $k$ qubits. Considering this, we obtain
Eq.~\eqref{eq:Nka-2}. The strong dependence of the bounds on $k$
in Eq.~\eqref{eq:Nka-2} indicates that for high precision metrology
states containing many-partite entanglement are needed. \ensuremath{\hfill \Box}

\textbf{\emph{Proof of Observation 4.}} Let us consider pure states
that are the tensor product of at most $k$-qubit entangled states
of the form Eq.~\eqref{eq:ttt-2} Hence, based on using Eq.~\eqref{eq:all}
for the $k$-qubit units, we obtain

\begin{eqnarray}
\sum_{l=x,y,z}(\Delta J_{l})_{\vert\Psi_{k-{\rm producible}}\rangle}^{2} & = & \sum_{m}\sum_{l=x,y,z}(\Delta J_{l})_{\vert\Psi_{m}^{(N_{m})}\rangle}^{2}\nonumber \\
 & \le & \sum_{m}\frac{N_{m}(N_{m}+2)}{4},
\end{eqnarray}
For the case when $k$ is a divisor of $N,$ the largest variance
can be obtained for a state for which all $N_{m}=k.$ Hence, for the
state Eq.~\eqref{eq:ttt-2} we obtain 
\begin{eqnarray}
\sum_{l=x,y,z}(\Delta J_{l})^{2} & \le & \frac{N}{k}\times\frac{k(k+2)}{4}.\label{eq:kk2}
\end{eqnarray}
If $k$ is not a divisor of $N,$ then at least one of the states
in the tensor product of Eq.~\eqref{eq:ttt-2} will have fewer than
$k$ qubits. The maximum for the sum of the variances is obtained
if all but a single state has $k$ qubits. Considering this, we obtain
Eq.~\eqref{eq:Nka}. We have to use that for pure states of $N\ge2$
qubits, we have $\sum_{k}(\Delta J_{l})^{2}\le\frac{N(N+2)}{4}$,
while for $N=1$ we have a better bound $\sum_{k}(\Delta J_{l})^{2}\le\frac{1}{2}.$
\ensuremath{\hfill \Box}

\textbf{\emph{Bound for states with a given number unentangled particles.}}
Next, we obtain bound for systems that contain a given number of unentangled
particles. A pure state is told to contain $M$ unentangled particles,
if it can be written as \cite{TK09,GT05}

\textbf{\emph{
\begin{equation}
\otimes_{k=1}^{M}\vert\Psi_{k}\rangle\otimes\vert\Psi_{M+1,..,N}\rangle.\label{eq:Mun}
\end{equation}
}}We say that a mixed state contains at least $M$ unentangled particles,
if it can be prepared by mixing pure states with $M$ or more unentangled
particles. 

Many interesting quantum states are highly entangled, but still contain
only two-particle entanglement. Nevertheless, it is still important
to know how large fraction of the particles remain unentangled since
the number of unentangled particles is directly connected to metrological
usefulness of the state. 

\textbf{\emph{Observation 5.}} For states with at least $M$ unentangled
particles, the quantum Fisher information is bounded from above by

\begin{eqnarray}
\sum_{l=x,y,z}F_{Q}[\varrho,J_{l}] & \le & M+(N-M)(N-M+2).\label{eq:F2ff}
\end{eqnarray}

\emph{Proof.} For a pure state of the form Eq.~\eqref{eq:Mun}, we
have 
\begin{eqnarray}
\sum_{l=x,y,z}(\Delta J_{l})^{2} & \le & \frac{M}{4}+\frac{(N-M)(N-M+2)}{4}.
\end{eqnarray}
Any state that violates Eq.~\eqref{eq:F2ff} has fewer than $M$
unentangled particles. The validity of Eq.~\eqref{eq:F2ff} for mixed
states is due to the convexity of the quantum Fisher information.
\ensuremath{\hfill \Box}

So far, we presented entanglement conditions in terms of $F_{Q}[\varrho,J_{l}]$
for $l=x,y,z.$ A more general approach is constructing entanglement
conditions with the $\Gamma_{C}$ matrix defined in Eq.~\ref{eq:GAMMAC}.
In Appendix A, we present unified framework for determining entanglement
conditions for $\Gamma_{C}$.

\section{Interesting points in the
$(F_{Q}[\varrho,J_{x}],F_{Q}[\varrho,J_{y}],F_{Q}[\varrho,J_{z}])$-space}

In this section, we discuss which part of the $(F_{Q}[\varrho,J_{x}],F_{Q}[\varrho,J_{y}],F_{Q}[\varrho,J_{z}])$-space
contains points corresponding to states with different degree of entanglement.
This is important, since apart from finding inequalities for states
of various types of entanglement, we have to show that there are states
that fulfill these inequalities. 

For that, let us see first the interesting points of the $(F_{Q}[\varrho,J_{x}],F_{Q}[\varrho,J_{y}],F_{Q}[\varrho,J_{z}])$-space
and the corresponding quantum states, shown in Fig.~\ref{fig:Points}.

(i) A completely mixed state 
\begin{equation}
\varrho_{C}=\frac{\openone}{2^{N}}.
\end{equation}
corresponds to the point $C(0,0,0)$ in the $(F_{Q}[\varrho,J_{x}],F_{Q}[\varrho,J_{y}],F_{Q}[\varrho,J_{z}])$-space.

(ii) Product states of the form 
\begin{equation}
\vert\Psi\rangle_{S_{l}}=\vert+\frac{1}{2}\rangle_{l}^{\otimes N/2}\otimes\vert-\frac{1}{2}\rangle_{l}^{\otimes N/2}
\end{equation}
for $l=x,y,z$ correspond to the points $S_{x}(0,N,N),S_{y}(N,0,N),$
and S$_{z}(0,N,N),$ respectively. 

(iii) An $N$-qubit symmetric Dicke state with $\frac{N}{2}$ excitations
in the $z$ basis is defined as
\begin{equation}
\vert\mathcal{D}_{N}^{(N/2)}\rangle=\binom{N}{N/2}^{-\frac{1}{2}}\sum_{k}\mathcal{P}_{k}\{\vert0)^{\otimes\frac{N}{2}}\otimes\vert1)^{\otimes\frac{N}{2}}\},
\end{equation}
where $\sum_{k}\mathcal{P}_{k}$ denotes summation over all possible
permutations. Such a state corresponds to the point $D_{z}(\frac{N(N+2)}{2},\frac{N(N+2)}{2},0).$
Dicke states in the $x$ and $y$ bases correspond to the points $D_{x}(0,\frac{N(N+2)}{2},\frac{N(N+2)}{2})$
and $D_{y}(\frac{N(N+2)}{2},0,\frac{N(N+2)}{2})$, respectively.

(iv) An $N$-qubit GHZ states in the $z$ basis is defined as 
\begin{equation}
\vert\Psi\rangle_{GHZ_{z}}=\frac{1}{\sqrt{2}}\left(\vert0\rangle^{\otimes N}+\vert1\rangle^{\otimes N}\right).
\end{equation}
It corresponds to the point $(N,N,N^{2}).$ GHZ states in the $x$
and $y$ bases correspond to points $(N^{2},N,N)$ and $(N,N^{2},N)$,
respectively. 

(v) Finally, the tensor product of a single-qubit state and a Dicke
state of the form 
\begin{equation}
\vert\Psi\rangle_{G_{Z}}=\vert1\rangle\otimes\vert\mathcal{D}_{N-1}^{(N/2-1)}\rangle.
\end{equation}
corresponds to the point $G_{z}(\frac{N^{2}}{2}+\frac{1}{2},\frac{N^{2}}{2}+\frac{1}{2},0)$
\cite{dickestates}. States corresponding to the points $G_{x}$ and
$G_{y}$ can be obtained from $\vert\Psi\rangle_{G_{Z}}$ by basis
transformations.

After considering individual points, we now show that there are two-dimensional
objects in the $(F_{Q}[\varrho,J_{x}],F_{Q}[\varrho,J_{y}],F_{Q}[\varrho,J_{z}])$-space
such that for all of their points there is a corresponding separable
or entangled quantum state.

(vi) For all points in the $S_{x},S_{y},S_{z}$ polytope, there is
a corresponding pure product state for even $N.$ Given $F[\varrho,J_{l}]$
for $l=x,y,z$, such a state is defined as
\begin{equation}
\varrho=\left[\frac{\openone}{2}+\frac{1}{2}\sum_{l=x,y,z}c_{l}\sigma_{l}\right]^{\otimes N/2}\otimes\left[\frac{\openone}{2}-\frac{1}{2}\sum_{l=x,y,z}c_{l}\sigma_{l}\right]^{\otimes N/2},
\end{equation}
where $c_{l}^{2}=1-\frac{F_{Q}[\varrho,J_{l}]}{N},$ where $\sum_{l}c_{l}^{2}=1.$ 

\begin{figure}
\begin{centering}
\includegraphics[width=8cm]{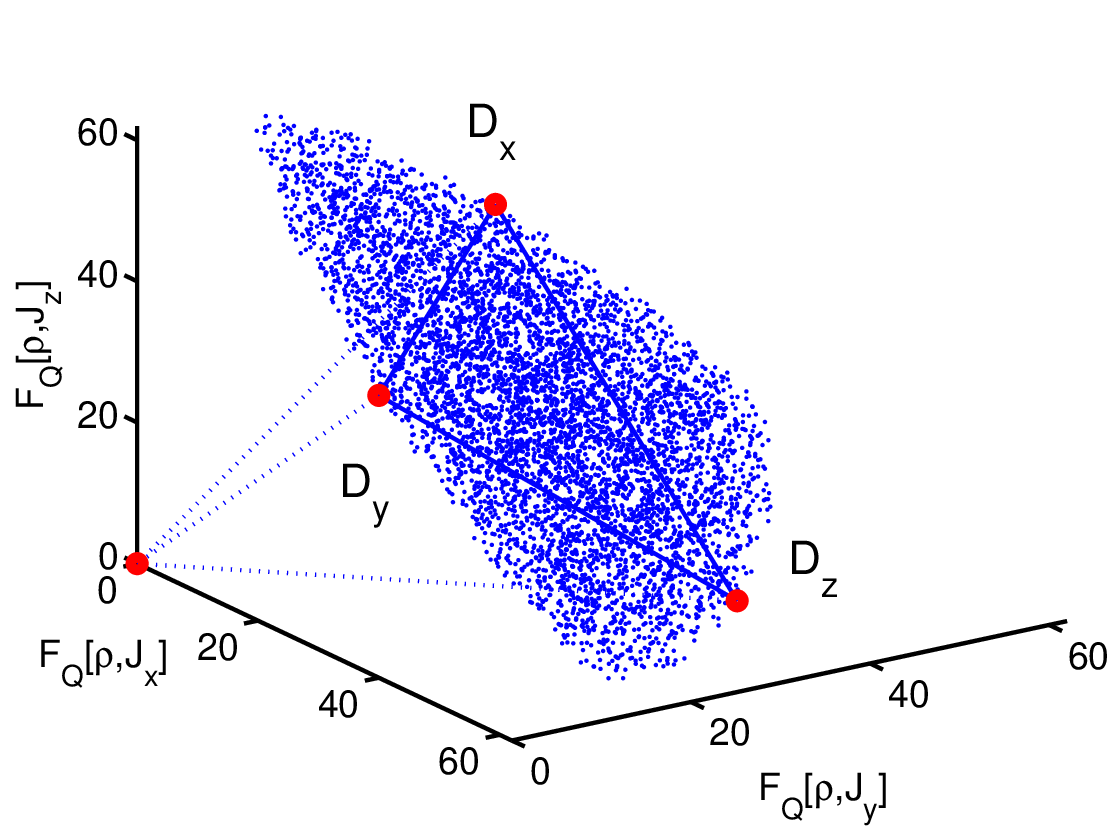}
\par\end{centering}

\caption{(Color online) Randomly chosen points in the $(F_{Q}[\varrho,J_{x}],F_{Q}[\varrho,J_{y}],F_{Q}[\varrho,J_{z}])$-space
corresponding to states of the form Eq.~\eqref{eq:psiprime2} for
$N=8.$ All the points are in the plane of $D_{x},D_{y}$ and $D_{z}.$
\label{fig:triangle}}
\end{figure}

(vii) For all points in the $D_{x},D_{y},D_{z}$ polytope, there is
a corresponding quantum state if $N$ is divisible by $4.$ To see
this, let us consider the following quantum states for even $N$
\begin{align}
\vert\Psi_{{\rm even}}\rangle & =\sum_{n=0,2,4,...,N/2-2}c_{n}\frac{1}{\sqrt{2}}\left(\vert\mathcal{D}_{N}^{(n)}\rangle+\vert\mathcal{D}_{N}^{(N-n)}\rangle\right)\nonumber \\
 & +c_{N/2}\vert\mathcal{D}_{N}^{(N/2)}\rangle,\label{eq:psiprime}
\end{align}
where $c_{n}$ are complex coefficients. States Eq.~\eqref{eq:psiprime}
are special cases of symmetric states with an even parity \cite{MW09}.
For $\vert\Psi_{{\rm even}}\rangle,$ we have $\langle J_{l}\rangle=0$
for $l=x,y,z.$ Finally, $\langle J_{l}J_{m}+J_{m}J_{l}\rangle=0$
if $l\ne m,$ thus for $\vert\Psi_{{\rm even}}\rangle$ the matrix
$\Gamma_{C}$ is diagonal. Let us now assume that $N$ is a multiple
of $4$ and consider can consider the states of the form
\begin{align}
\vert\Psi(\alpha_{x},\alpha_{y},\alpha_{z})\rangle\;\,\,\,\,\;\;\;\;\;\;\;\;\;\;\;\,\,\,\,\;\;\;\;\;\;\;\;\;\;\;\;\;\;\;\;\;\;\;\;\;\;\;\;\;\;\;\;\;\;\;\;\;\;\;\;\;\;\;\;\;\nonumber \\
=\alpha_{x}\vert\mathcal{D}_{N}^{(N/2)}\rangle_{x}+\alpha_{y}\vert\mathcal{D}_{N}^{(N/2)}\rangle_{y}+\alpha_{z}\vert\mathcal{D}_{N}^{(N/2)}\rangle_{z},\label{eq:psiprime2}
\end{align}
where $\alpha_{l}$ are complex coefficients. (Note that $\vert\mathcal{D}_{N}^{(N/2)}\rangle_{l}$
are not pairwise orthogonal.) Simple analytical calculations show
that such states are a subset of the states Eq.~\eqref{eq:psiprime}
\cite{evenstates}. The states \eqref{eq:psiprime2} fill the polytope
$D_{x},$ $D_{y},$ and $D_{z},$ which is demonstrated for $N=8$
in Fig.~\ref{fig:triangle} \cite{QUBIT4MATLAB} (see also Appendix
B). Thus, there is a quantum state corresponding to all points of
this polytope. 

Next, we examine, how to obtain states corresponding to three-dimensional
polytopes. For that we  use that when mixing two states, the points
corresponding to the mixed state are on a curve in the $(F_{Q}[\varrho,J_{x}],F_{Q}[\varrho,J_{y}],F_{Q}[\varrho,J_{z}])$-space.
In the general case, this curve is not a straight line. For the case
of mixing a pure state with the completely mixed state, the curve
is a straight line. Such a state is defined as

\begin{equation}
\varrho^{({\rm mixed})}(p)=p\varrho+(1-p)\frac{\openone}{2^{N}}.
\end{equation}
Using Eq.~\eqref{eq:GAMMAC}, after simple calculations we have
\begin{equation}
\Gamma_{C}^{({\rm mixed})}(p)=\frac{p^{2}}{p+(1-p)2^{-(N-1)}}\Gamma_{C}^{(\varrho)}.
\end{equation}

Hence we can state the following. 

\textbf{Observation 6. }If $N$ is even then there is a separable
state for each point in the $S_{x},S_{y},S_{z},C$ polytope\emph{.}

\emph{Proof.} This is because there is a pure product state corresponding
to any point in the $S_{x},S_{y},S_{z}$ polytope. When mixing any
of these states with the completely mixed state, we obtain states
that correspond to points on the line connecting the pure state to
point C. \ensuremath{\hfill \Box}

\textbf{Observation 7.} If $N$ is divisible by $4,$ then for all
the points of the $D_{x},D_{y},D_{z},G_{x},G_{y},G_{z}$ polytope,
there is a quantum state with genuine multipartite entanglement.

\emph{Proof.} There is a quantum state for all points in the $D_{x},D_{y},D_{z}$
polytope. Mixing them with the completely mixed state, states corresponding
to all points of the $C,D_{x},D_{y},D_{z}$ polytope can be obtained.
Based on Observation 2, states corresponding to the points in the
$D_{x},D_{y},D_{z},G_{x},G_{y},G_{z}$ polytope are genuine multipartite
entangled. \ensuremath{\hfill \Box}

Finally, note that all the quantum states we presented in this section
have a diagonal $\Gamma_{C}$ matrix. Thus, our statements remain
true even if the three coordinate axes in Fig.~\ref{fig:Points}
correspond to the three eigenvalues of $\Gamma_{C}$.

\section{Discussion}

The criterion in Eq.~\eqref{eq:F2ea} contains several quantum Fisher
information terms. It can happen that a state does not violate the
criterion Eq.~\eqref{eq:F2eb}, but it violates the criterion Eq.~\eqref{eq:F2ea}.
In this case, for a single metrological task of the type we considered
in this paper its entanglement does not make it possible to outperform
the metrology with separable states. However, if the state is used
for several metrological tasks, then it makes it possible to achieve
such an \emph{average} sensitivity that would be not possible for
separable states. 

A related example is the proposal of using multipartite singlets for
differential magnetometry \cite{TM10}. Singlets are useful for differential
magnetometry because they are insensitive to homogeneous fields, that
is, $F[\varrho,J_{l}]=0$ for $l=x,y,z,$ which is the same as for
the completely mixed state. However, when considering operators other
than $J_{\vec{n}},$ singlets turn out to be very sensitive, which
is not the case for the completely mixed state. Thus, singlets can
provide an advantage over separable states if the combination of two
metrological tasks are considered. 

It is instructive to compare the necessary condition for separability
Eq.~\eqref{eq:F2ea} to the condition presented in Ref.~\cite{T04,WV05}
\begin{equation}
\sum_{l=x,y,z}(\Delta J_{l})^{2}\ge\frac{N}{2}.\label{eq:sumofvariances}
\end{equation}
Clearly, if a pure state is detected by Eq.~\eqref{eq:sumofvariances},
it is not detected by Eq.~\eqref{eq:F2ea}, and vice versa. In fact,
Eqs.~\eqref{eq:sumofvariances} and \eqref{eq:F2ea} together detect
all entangled pure multiqubit states except for the ones for which 

\begin{equation}
\sum_{l=x,y,z}(\Delta J_{l})^{2}=\frac{N}{2}.
\end{equation}
Of course, the two conditions also detect some mixed entangled states
in the vicinity of the pure entangled states. 

It is an interesting question whether multipartite states having a
positive partial transpose for all bipartitions can violate any of
the above entanglement criteria with the quantum Fisher information.
Violating Eq.~\eqref{eq:F2ea} would certainly mean that such bound
entangled states are useful for certain metrological applications.
To find such states, if they exist, might be difficult as typically
bound entangled states are strongly mixed and the quantum Fisher information
is convex. 

Concerning multipartite entanglement, Observation 3 shows that for
a single metrological task, genuine multipartite entanglement is needed
to reach the maximum sensitivity. Observation 4 demonstrates that
even for the maximum average sensitivity for the metrological tasks
considered can be reached only by states possessing genuine multipartite
entanglement.

Finally, the definition of quantum Fisher information used in Ref.
\cite{PS01}, while widely considered {}``the'' quantum Fisher information,
is not the only possible definition \cite{PRIV}. The Wigner-Yanase
skew information is another possibility \cite{WY63,P02,L03}. This
quantity equals to the variance for pure states, and it is also convex
in the state. This has already been used to define entanglement criteria
with the skew information \cite{M09,C05}. Thus, all previous statements
can easily be transformed into criteria with the skew information.

\section{Conclusions}

In summary, we showed that genuine multipartite entanglement, or in
general, multipartite entanglement more demanding than simple inseparability,
is needed to achieve a maximal accuracy using multipartite quantum
states for metrology. We also considered several relations with the
quantum Fisher information and determined the corresponding bounds
for various forms of entanglement. 

\emph{Note added in proof. }Independently from our work, a paper on
the relationship between multipartite entanglement and Fisher information
has been prepared \cite{HS00}.
\begin{acknowledgments}
We thank O.~G\"uhne and D.~Petz for discussions. We thank the European
Union (ERC Starting Grant GEDENTQOPT and CHIST-ERA QUASAR), the Spanish
MICINN (Project No. FIS2009-12773-C02-02), the Basque Government (Project
No. IT4720-10), and the support of the National Research Fund of Hungary
OTKA (Contract No. K83858).
\end{acknowledgments}
\appendix

\section{Entanglement conditions for the \textbf{\emph{$\Gamma_{C}$}} matrix}

In this appendix, we present a unified framework to derive entanglement
conditions for the \textbf{\emph{$\Gamma_{C}$}} matrix. For that
aim, we use ideas from the derivation of the Covariance Matrix Criterion
\cite{G04,GH07} and the entanglement criteria for Gaussian multi-mode
states \cite{DG00,S00}. We recall that a separable state is a mixture
of pure product states \cite{W89}

\begin{equation}
\varrho_{{\rm sep}}=\sum_{k}p_{k}\rho_{{\rm pure\, product},k}.
\end{equation}
Due to the convexity of the quantum Fisher information \cite{PS01},
we have 

\begin{equation}
F[\varrho_{{\rm sep}},J_{\vec{n}}]\le\sum_{k}p_{k}F[\rho_{{\rm pure\, product},k},J_{\vec{n}}].\label{eq:prod}
\end{equation}
Thus, for every separable state there must be a set of $p_{k}$ and
$\rho_{{\rm pure\, product},k}$ fulfilling Eq.~\eqref{eq:prod}.
Hence, we can say the following. For every separable state, there
is a set of $p_{k}$ and $\rho_{{\rm pure\, product},k}$ such that

\begin{equation}
\Gamma_{C}^{({\rm sep})}\le\sum_{k}p_{k}\text{\text{\ensuremath{\Gamma_{C}^{({\rm {\rm pure\,\ product}},k)}}}}.\label{eq:Gamma}
\end{equation}
Any state for which there are not such a set of probabilities and
pure product density matrices is entangled \cite{Remark}.

It is known that for $N$-qubit pure product states we have the following
two constraints for the variances of the three angular momentum components
\begin{subequations}\label{eq:lambdamaxx-1}

\begin{eqnarray}
\sum_{l=x,y,z}(\Delta J_{l})^{2} & = & \frac{N}{2},\label{eq:lambdamaxx-2}\\
(\Delta J_{m})^{2} & \le & \frac{N}{4}.\label{eq:qqww}
\end{eqnarray}
\end{subequations} which has been used to derive entanglement criteria
with the three variances \cite{T04,WV05,TK07,TK09}. Equation~\eqref{eq:lambdamaxx-2}
also appeared in the proof of Observation 1. Based on Eqs.~\eqref{eq:lambdamaxx-2}
and \eqref{eq:qqww}, the conditions for the eigenvalues of $\Gamma_{C}^{({\rm {\rm pure\, product}})}$
are clearly

\begin{eqnarray}
\sum_{l=x,y,z}\Lambda_{l}^{({\rm pure\, product})} & = & 2N,\nonumber \\
0\le\Lambda_{m}^{({\rm pure\, product})} & \le & N,\label{eq:F2e33}
\end{eqnarray}
for $m=x,y,z.$ Using now our knowledge about $\text{\ensuremath{\Gamma_{C}^{({\rm {\rm pure\,\ product}},k)}},}$
the condition Eq.~\eqref{eq:Gamma} leads to the following equations
for the eigenvalues of $\Gamma_{C}^{({\rm sep})}$ \begin{subequations}\label{eq:lambdamaxx}

\begin{eqnarray}
\sum_{l=x,y,z}\Lambda_{l}^{({\rm sep})} & \le & 2N,\label{eq:F2e33yya}\\
0\le\Lambda_{m}^{({\rm sep})} & \le & N,\label{eq:F2e33yy}
\end{eqnarray}
\end{subequations} for $m=x,y,z.$ Equation~\eqref{eq:lambdamaxx}
can be reformulated with $\Gamma_{C}$ as \begin{subequations}\label{eq:lambdamax}

\begin{eqnarray}
{\rm Tr}(\Gamma_{C}^{({\rm sep})}) & \le & 2N,\label{eq:lambdamaxa}\\
\Lambda_{{\rm max}}(\Gamma_{C}^{({\rm sep})}) & \le & N,\label{eq:lambdamaxb}
\end{eqnarray}
\end{subequations} where $\Lambda_{{\rm \max}}(A)$ is the largest
eigenvalue of $A.$ Equation~\eqref{eq:lambdamaxb} has appeared
in Ref.~\cite{HG09}. 

Hence, quantum states fulfilling Eq.~\eqref{eq:Gamma} must fulfill
Eq.~\eqref{eq:lambdamax}. In Observation 1 and also for the criterion
Eq.~\eqref{eq:F2eb}, the most entangled states are detected if $F[\varrho_{{\rm sep}},J_{l}]$
correspond to the three eigenvalues of $\Gamma_{C}.$ For this case
Eq.~\eqref{eq:lambdamaxa} is equivalent to Observation 1 and Eq.~\eqref{eq:lambdamaxb}
is equivalent to Eq.~\eqref{eq:F2eb}.

In a similar manner, conditions for multipartite entanglement can
also be obtained. Thus, analogously to Observations 3 and Observation
4, for $N$-qubit $k$-producible states states, we obtain \begin{subequations}
\begin{align}
{\rm Tr}(\Gamma_{C}^{({\rm sep})})\le\;\;\;\;\;\;\;\;\;\;\;\;\;\;\;\;\;\;\;\;\;\;\;\;\;\;\;\;\;\;\;\;\;\;\;\;\;\;\;\;\;\;\;\;\;\;\;\;\;\;\;\;\;\;\;\;\;\;\;\;\;\;\;\;\nonumber \\
\begin{cases}
\begin{array}{c}
nk(k+2)+(N-nk)(N-nk+2)\\
nk(k+2)+2
\end{array} & \begin{array}{c}
\text{if }N-nk\ne1,\\
\text{if }N-nk=1,
\end{array}\end{cases}\label{eq:gme_a}\\
\Lambda_{{\rm max}}(\Gamma_{C}^{({\rm sep})})\le nk^{2}+(N-nk)^{2},\,\,\,\,\,\,\,\,\,\,\,\,\,\,\,\,\,\,\,\,\,\,\,\,\,\,\,\,\,\,\,\,\,\,\,\,\,\,\,\,\,\,\,\,\,\,\,\,\,\,\,\,\,\,\,\,\,\,\,\label{eq:gme_b}
\end{align}
\end{subequations} where $n$ is the largest integer such that $nk\le N.$
We can obtain the bounds for biseparability setting $n=1$ and $k=N-1.$
Any state that violates one of the criteria for $n=1$ and $k=N-1,$
is genuine multipartite entangled. The inequalities \eqref{eq:gme_a}
and \eqref{eq:gme_b} are essentially the criteria of Observation
3 and 4 rewritten in a coordinate system independent way.

\section{$\Gamma_C$ matrix for the state Eq.~(\ref{eq:psiprime2})}

In this appendix, we compute the $\Gamma_{C}$ matrix for the superposition
of three Dicke states given in Eq.~\eqref{eq:psiprime2}. We show
that for any point in the $D_{x},$ $D_{y},$ $D_{z}$ triangle in
Fig.~\ref{fig:triangle} there is a corresponding state of this type.

First, we need to know that
\begin{align}
 & _{k}\langle\mathcal{D}_{N}^{(N/2)}\vert J_{l}^{2}\vert\mathcal{D}_{N}^{(N/2)}\rangle_{m}\nonumber \\
 & \,\,\,\,=\begin{cases}
\begin{array}{ll}
\frac{N(N+2)}{8} & {\rm \text{if }}k=m\ne l,\\
Q & {\rm \text{if }}k\ne m{\rm \text{ and }}m\ne l\text{ and }k\ne l,\\
0 & {\rm \text{otherwise},}
\end{array}\end{cases}\label{eq:klm}
\end{align}
for $k,l,m\in\{x,y,z\}.$ In the second line on the right hand side
of Eq.~\eqref{eq:klm}, $Q=_{x}\langle\mathcal{D}_{N}^{(N/2)}\vert J_{y}^{2}\vert\mathcal{D}_{N}^{(N/2)}\rangle_{z}.$
Since the state vector of $\vert\mathcal{D}_{N}^{(N/2)}\rangle_{x}$
and$\vert\mathcal{D}_{N}^{(N/2)}\rangle_{z}$ all have real elements,
and $J{}_{y}^{2}$ also have only real elements for even $N,$ $Q$
is also real. Its precise value is not important for proving the main
statement of this section. The last line on the right hand side of
Eq.~\eqref{eq:klm} is due to the fact that $J_{l}\vert\mathcal{D}_{N}^{(N/2)}\rangle_{l}=0.$ 

Hence, the $\Gamma_{C}$ matrix for state Eq.~\eqref{eq:psiprime2}
is a diagonal matrix, with 
\begin{equation}
\Gamma_{C,xx}=(\vert\alpha_{y}\vert^{2}+\vert\alpha_{z}\vert^{2})\frac{N(N+2)}{2}+2\Re(\alpha_{y}^{*}\alpha_{z}Q).
\end{equation}
The elements $\Gamma_{C,yy}$ and $\Gamma_{C,zz}$ can be obtained
in a similar way, after relabeling the coordinates. Clearly, for $(\alpha_{x},\alpha_{y},\alpha_{z})=(1,0,0)$,
the state Eq.~\eqref{eq:psiprime2} corresponds to the $D_{x}$ point
in Fig.~\ref{fig:triangle}. Similarly, $(\alpha_{x},\alpha_{y},\alpha_{z})=(0,1,0$)
and $(0,0,1)$ correspond to the $D_{y}$ and $D_{z}$ points, respectively.
With an appropriate choice of phases for $\alpha_{i},$ a state with
$\vert\alpha_{x}\vert=\vert\alpha_{y}\vert=\vert\alpha_{z}\vert$
correspond to the center of the $D_{x},D_{y},D_{z}$ triangle. Moreover,
a state with $\alpha_{x}=i\alpha_{y}$ and $\alpha_{z}=0$ corresponds
to a point halfway between $D_{x}$ and $D_{y}.$ In a similar manner,
states of the form Eq.~\eqref{eq:psiprime2} can be obtained for
the points halfway between $D_{x}$ and $D_{z},$ and $D_{y}$ and
$D_{z}.$

Based on similar arguments, one can show that with the appropriate
choice of the absolute values and phases of $\alpha_{k},$ it is possible
to get all the matrices 

\begin{align}
\Gamma_{c} & =\alpha_{x}'{\rm diag}\left(0,\frac{N(N+2)}{2},\frac{N(N+2)}{2}\right)\nonumber \\
 & +\alpha_{y}'{\rm diag}\left(\frac{N(N+2)}{2},0,\frac{N(N+2)}{2}\right)\nonumber \\
 & +\alpha'_{z}{\rm diag}\left(\frac{N(N+2)}{2},\frac{N(N+2)}{2},0\right)
\end{align}
with $0\le\alpha_{l}'\le1$ and $\alpha_{x}'+\alpha_{y}'+\alpha_{z}'=1.$
That is, we can get any point corresponding of the $D_{x},$ $D_{y},$
$D_{z}$ triangle in Fig.~\ref{fig:triangle}.

$ $


\begin{thebibliography}{References}
\bibitem{PB00}J.-W. Pan, D. Bouwmeester, M. Daniell, H. Weinfurter,
and A. Zeilinger, Nature (London) \textbf{403}, 515 (2000).

\bibitem{BE04}M. Bourennane, M. Eibl, C. Kurtsiefer, S. Gaertner,
H. Weinfurter, O. G\"uhne, P. Hyllus, D. Bru\ss, M. Lewenstein,
and A. Sanpera, Phys. Rev. Lett. \textbf{92}, 087902 (2004). 

\bibitem{KS05}N. Kiesel, C. Schmid, U. Weber, G. T\'oth, O. G\"uhne,
R. Ursin, and H. Weinfurter, Phys. Rev. Lett. \textbf{95}, 210502
(2005).

\bibitem{KS07}N. Kiesel, C. Schmid, G. T\'oth, E. Solano, and H.
Weinfurter, Phys. Rev. Lett. \textbf{98}, 063604 (2007).

\bibitem{WK09}W. Wieczorek, R. Krischek, N. Kiesel, P. Michelberger,
G. T\'oth, and H. Weinfurter, Phys. Rev. Lett. \textbf{103}, 020504
(2009); G. T\'oth, W. Wieczorek, R. Krischek, N. Kiesel, P. Michelberger,
and H. Weinfurter, New J. Phys. \textbf{11}, 083002 (2009).

\bibitem{PC09}R. Prevedel, G. Cronenberg, M. S. Tame, M. Paternostro,
P. Walther, M. S. Kim, A. Zeilinger, Phys. Rev. Lett. \textbf{103},
020503 (2009); S. Campbell, M. S. Tame, M. Paternostro, New J. Phys.
\textbf{11,} 073039 (2009).

\bibitem{SK00}C. A. Sackett, D. Kielpinski, B. E. King, C. Langer,
V. Meyer, C. J. Myatt, M. Rowe, Q. A. Turchette, W. M. Itano, D. J.
Wineland, and C. Monroe, Nature (London) \textbf{404}, 256 (2000).

\bibitem{HH05}H. H\"affner, W. H\"ansel, C. Roos, J. Benhelm, D.
Chek-al-Kar, M. Chwalla, T. K\"orber, U. D. Rapol, M. Riebe, P. O.
Schmidt, C. Becher, O. G\"uhne, W. D\"ur, R. Blatt, Nature (London)
\textbf{438}, 643, (2005).

\bibitem{MG03}O. Mandel, M. Greiner, A. Widera, T. Rom, T.W. Hänsch,
and I. Bloch, Nature (London) \textbf{425}, 937 (2003).

\bibitem{AB01}A. Ac\'{\i}n, D. Bru\ss, M. Lewenstein, and A. Sanpera,
Phys. Rev. Lett. \textbf{87}, 040401 (2001).

\bibitem{GT09}O. G\"uhne and G. T\'oth, Phys. Rep. \textbf{474},
1 (2009).

\bibitem{GS04}V. Giovannetti, S. Lloyd, and L. Maccone, Science \textbf{306},
1330 (2004).

\bibitem{H82}A. S. Holevo, \emph{Probabilistic and Statistical Aspect
of Quantum Theory} (North-Holland, Amsterdam, 1982).

\bibitem{H76}C.W. Helstrom, \emph{Quantum Detection and Estimation
Theory} (Academic Press, New York, 1976). 

\bibitem{PS01}L. Pezz\'e and A. Smerzi, Phys. Rev. Lett. \textbf{102},
100401 (2009). 

\bibitem{HG09}P. Hyllus, O. G\"uhne, and A. Smerzi, Phys. Rev. A
\textbf{82}, 012337 (2010).

\bibitem{SM01}A.S. S\o rensen and K. M\o lmer, Phys. Rev. Lett.
\textbf{86}, 4431 (2001).

\bibitem{EntWit}For entanglement witnesses, see M. Horodecki, P. Horodecki, and R. Horodecki, Phys. Lett. A {\bf 223}, 1 (1996); B.M. Terhal, Phys. Lett. A {\bf 271}, 319 (2000); M. Lewenstein, B. Kraus, J.I. Cirac, and P. Horodecki, Phys. Rev. A {\bf 62}, 052310 (2000); D. Bru{\ss}, J.I. Cirac, P. Horodecki, F. Hulpke, B. Kraus, M. Lewenstein, and A. Sanpera, J. Mod. Opt. {\bf 49}, 1399 (2002). For the detection of genuine multipartite entanglement, see M. Bourennane, M. Eibl, C. Kurtsiefer, S. Gaertner, H. Weinfurter, O. G\"uhne, P. Hyllus, D. Bru{\ss}, M. Lewenstein, and A. Sanpera, Phys. Rev. Lett. {\bf 92}, 087902 (2004);  G. T\'oth and O. G\"uhne, Phys. Rev. Lett. {\bf 94}, 060501 (2005); G.A. Durkin and C. Simon, Phys. Rev. Lett. {\bf 95}, 180402 (2005).

\bibitem{TG10}It has also been worked out how to detect the genuine
multipartite entanglement that can be obtained in a selected part
of a very large quantum system through local operations around the
boundary of that selected part. This makes it possible to study multipartite
entanglement in the three-, four-, and five-particle blocks of a large
quantum system and produce an entanglement map. See E. Alba, G. T\'oth, and J.J. Garc\'{\i}a-Ripoll, Phys. Rev. A {\bf 82}, 062321 (2010).

\bibitem{JM11}Recently, via semidefinite programming, it has become
possible to find an entanglement witness detecting genuine multipartite
entanglement for a given quantum state. See B. Jungnitsch, T. Moroder,
and O. G\"uhne, Phys. Rev. Lett. \textbf{106}, 190502 (2011).

\bibitem{BG11}For device independent entanglement witnesses for multipartite
entanglement see J.-D. Bancal, N. Gisin, Y.-C. Liang, and S. Pironio,
Phys. Rev. Lett. \textbf{106}, 250404 (2011).

\bibitem{B64}J.S. Bell, Physics (Long Island City, N.Y.) \textbf{1},
195 (1964). 

\bibitem{M90}N.D. Mermin, Phys. Rev. Lett. \textbf{65}, 1838 (1990).

\bibitem{GB98}N. Gisin and H. Bechmann-Pasquinucci, Phys. Lett A
\textbf{246}, 1 (1998). 

\bibitem{SU01}M. Seevinck and J. Uffink, Phys. Rev. A. \textbf{65},
012107 (2001). 

\bibitem{CN02}D. Collins, N. Gisin, S. Popescu, D. Roberts, and V.
Scarani, Phys. Rev. Lett. \textbf{88}, 170405 (2002).

\bibitem{NK02}K. Nagata, M. Koashi, and N. Imoto, Phys. Rev. Lett.
\textbf{89}, 260401 (2002).

\bibitem{U02}J. Uffink, Phys. Rev. Lett. \textbf{88}, 230406 (2002).

\bibitem{dVH11}J.I. de Vicente, and M. Huber, arXiv:1106.5756.

\bibitem{SG09}M. Seevinck and O. Gühne, New J. Phys \textbf{12},
053002 (2010). 

\bibitem{HM10}M. Huber, F. Mintert, A. Gabriel, and B.C. Hiesmayr,
Phys. Rev. Lett. \textbf{104}, 210501 (2010).

\bibitem{T07}G. T\'oth, J. Opt. Soc. Am. B \textbf{24}, 275 (2007).

\bibitem{VH11}G. Vitagliano, P. Hyllus, I.L. Egusquiza, and G. T\'oth,
Phys. Rev. Lett, in press; arXiv:1104.3147.

\bibitem{D11}L.-M. Duan, Phys. Rev. Lett. \textbf{107}, 180502 (2011).

\bibitem{C05}Z. Chen, Phys. Rev. A \textbf{71,} 052302 (2005).

\bibitem{GHZ}D. M. Greenberger, M. A. Horne, A. Shimony, and A. Zeilinger,
Am. J. Phys. \textbf{58}, 1131 (1990).

\bibitem{GT05}O. G\"uhne, G. T\'oth, and H.J. Briegel, New J. Phys.
\textbf{7}, 229 (2005).

\bibitem{PH}We thank P. Hyllus for pointing out that the $N-nk=1$
case is special.

\bibitem{BC94}S.L. Braunstein and C.M. Caves, Phys. Rev. Lett. \textbf{72},
3439 (1994).

\bibitem{TK09}G. T\'oth, C. Knapp, O. G\"uhne, and H.J. Briegel,
Phys. Rev. A \textbf{79}, 042334 (2009).

\bibitem{dickestates}For the values of $(\Delta J_{l})^{2}$ for
$l=x,y,z$ for Dicke states, see Eq.~(25) of Ref.~\cite{TK09}.

\bibitem{MW09}X. Yin, X. Wang, J. Ma, and X. Wang, J. Phys. B: At.
Mol. Opt. Phys. \textbf{44}, 015501 (2011).

\bibitem{evenstates} In Ref. \cite{MW09}, it has been shown that
for states with an even parity $\langle J_{z}J_{l}+J_{l}J_{z}\rangle=0$
for $l=x,y.$ For states of the form Eq.~\eqref{eq:psiprime}, $\langle J_{x}J_{y}+J_{y}J_{x}\rangle=0$
due to $\vert\Psi_{{\rm even}}\rangle=\sigma_{x}^{\otimes N}\vert\Psi_{{\rm even}}\rangle.$
Eq.~\eqref{eq:psiprime2} is of the form Eq.~\eqref{eq:psiprime}
because for this state $\vert\Psi(\alpha_{x},\alpha_{y},\alpha_{z})\rangle=\sigma_{x}^{\otimes N}\vert\Psi(\alpha_{x},\alpha_{y},\alpha_{z})\rangle,$
and the overlap of this state with symmetric Dicke states with an
odd number of $1$'s is zero, which can be seen as follows. When writing
$\vert D_{N}^{(N/2)}\rangle_{x}$ in the $x$ basis, we find that
it is an equal superposition of several computational basis states
in the $x$ basis. If $\vert b_{1}b_{2}...b_{N}\rangle_{x}$ appears
in this superposition, so does $\vert\overline{b}_{1}\overline{b}_{2}...\overline{b}_{N}\rangle_{x}$,
where $b\in\{0,1\}$ and $\overline{b}$ denotes the logical inversion.
All the terms of the superposition have $\frac{N}{2}$ $1$'s and
$\frac{N}{2}$ $0$'s. Based on this, after straightforward calculation
one finds that $\langle D_{N}^{(m)}\vert(\vert b_{1}b_{2}...b_{N}\rangle_{x}+\vert\overline{b}_{1}\overline{b}_{2}...\overline{b}_{N}\rangle_{x})=0$
for odd $m.$ Hence, $\langle D_{N}^{(m)}\vert D_{N}^{(N/2)}\rangle_{x}=0$
for odd $m$ follows. Similar calculation can be carried out for $\langle D_{N}^{(m)}\vert D_{N}^{(N/2)}\rangle_{y}.$

\bibitem{QUBIT4MATLAB}The calculations have been made with QUBIT4MATLAB
V3.0. See G. T\'oth, Comput. Phys. Comm. \textbf{179}, 430 (2008).

\bibitem{uncent}For the general theory of entanglement detection
with uncertainty relations, see H. F. Hofmann and S. Takeuchi, Phys.
Rev. A \textbf{68}, 032103 (2003); O. G\"uhne, Phys. Rev. Lett. \textbf{92},
117903 (2004). 

\bibitem{T04}G. T\'oth, Phys. Rev. A \textbf{69}, 052327 (2004).

\bibitem{WV05}M. Wie\' sniak, V. Vedral, and \v C. Brukner, New
J. Phys. \textbf{7}, 258 (2005).

\bibitem{TM10}G. Tóth and M.W. Mitchell, New J. Phys \textbf{\noun{12}},
053007 (2010).

\bibitem{PRIV}D. Petz, private communication (2009).

\bibitem{P02}D. Petz, J. Phys. A: Math. Gen. \textbf{35}, 929 (2002).

\bibitem{L03}S.-L. Luo, Phys. Rev. Lett. \textbf{91}, 180403 (2003).

\bibitem{WY63}E.P. Wigner and M.M. Yanase, Proc. Natl. Acad. Sci.
U.S.A. \textbf{49}, 910 (1963).

\bibitem{M09}Zh. Ma, arXiv:0908.1291.

\bibitem{HS00}P. Hyllus, W. Laskowski, R. Krischek, C. Schwemmer,
W. Wieczorek, H. Weinfurter, L. Pezz\'e, and A. Smerzi, preceding
paper Phys. Rev. A \textbf{85}, 022321 (2012); \href{http://arxiv.org/abs/1006.4366}{arxiv:1006.4366}.
See also R. Krischek, C. Schwemmer, W. Wieczorek, H. Weinfurter, P.
Hyllus, L. Pezz\'e, and A. Smerzi, Phys. Rev. Lett. \textbf{107},
080504 (2011).

\bibitem{G04}O. G\"uhne, Phys. Rev. Lett. \textbf{92}, 117903 (2004)
.

\bibitem{GH07}O. G\"uhne, P. Hyllus, O. Gittsovich, and J. Eisert,
Phys. Rev. Lett. \textbf{99}, 130504 (2007).

\bibitem{DG00}L.-M. Duan, G. Giedke, J.I. Cirac, and P. Zoller, Phys.
Rev. Lett. \textbf{84}, 2722 (2000).

\bibitem{S00}R. Simon, Phys. Rev. Lett. \textbf{84}, 2726 (2000).

\bibitem{W89}R. F. Werner, Phys. Rev. A \textbf{40}, 4277 (1989).

\bibitem{Remark}Note that this idea can also be applied for the covariance
matrix defined as $[\Gamma]_{ij}=\langle J_{i}J_{j}+J_{j}J_{i}\rangle/2-\langle J_{i}J_{j}\rangle.$
Due to the concavity of the variance, for any separable state there
must be set of $p_{k}$ and $\rho_{{\rm pure\, product},k}$ such
that $\Gamma^{({\rm sep})}\ge\sum_{k}p_{k}\text{\text{\ensuremath{\Gamma^{({\rm {\rm pure\,\ product}},k)}}}}.$

\bibitem{TK07}G. T\'oth, C. Knapp, O. G\"uhne, and H.J. Briegel,
Phys. Rev. Lett. \textbf{99}, 250405 (2007).\end{thebibliography}
\end{document}